\documentstyle[pra,aps,multicol,psfig]{revtex}

\def\addcontentsline#1#2#3{\relax}
\begin{document}
\draft
\title{Broken symmetry, hyper-fermions and universal conductance in transport through a fractional quantum Hall edge.}
\author{V.V. Ponomarenko  and D.V. Averin}
\address{Department of Physics and Astronomy, SUNY, Stony Brook, NY 11794, USA}
\date{\today}
\maketitle
\begin{abstract}
We have found solution to a model of tunneling between a multi-channel Fermi liquid reservoir and an edge of the principal fractional quantum Hall liquid (FQHL) in the strong coupling limit. The solution explains how the absence of the time-reversal symmetry at high energies due to chiral edge propagation makes the universal two-terminal conductance of the FQHL fractionally quantized and different from that of a 1D Tomonaga-Luttinger liquid wire, where a similar model but preserving the time-reversal symmetry predicts unsuppressed free-electron conductance.
\end{abstract}


%
\multicols{2}
Low energy transport through an incompressible Quantum Hall liquid (QHL) with gapped bulk excitations is carried by gapless edge modes \cite{edge,butt}. For principal FQHL of the filling factor $\nu=1/odd$ these modes are described as a single branch of a chiral Luttinger liquid ($\chi$LL) \cite{wen1,hyper}. In presence of the right and left chiral edges, the model of the FQHL transport \cite{wen2} appears to be equivalent to that of a metallic phase of a 1D interacting electron gas \cite{kf} known as the Tomonaga-Luttinger liquid (TLL). To describe the two-terminal transport experiments, external reservoirs have to be added \cite{land,butt} to the model so that the full transport process includes transformation of the reservoir electrons into the FQHL/TLL quasiparticles in the junctions. The transformation process makes the two-terminal conductance of both the TLL wire \cite{tar,me} and the narrow FQHL junctions \cite{chkl,chamon} equal to the free electron conductance $\sigma_0$. The standard experimental observation, however, is that the two-terminal FQHL conductance is equal to the Hall conductivity $\nu \sigma_0$ (see, e.g., \cite{chang}) but not $\sigma_0$, the fact that implies equilibration between the chemical potentials of the reservoirs and the outgoing edges \cite{wen3}. This problem was recently studied \cite{kfcontact,chamon} for the junction modeled as a sequence of point-like contacts between the edge and different channels of a multi-channel Fermi-liquid (FL) reservoir under the  assumption of suppressed quantum interference between electron tunneling at different  contacts \cite{chamon}, e.g., due to finite decoherence length \cite{kfcontact}.  Then, currents injected into the edge at different points are directly related to the local values of the chemical potentials and can be added incoherently into a common current-voltage characteristic. Although this procedure reproduces equilibration between the edge and reservoir as  the number of reservoir channels is increased, the model does not provide the reason for suppression of  quantum interference of the channels at low energies. This drawback is very important, since  in the low-energy limit any finite distance between the channels drops out from the solution and quantum interference is known \cite{tar,me} to reduce all the channels to a single tunneling mode. Therefore a quantum solution is needed to determine whether the multi-channel tunneling model is complete or requires an additional decoherence mechanism for equilibration at physically relevant low energies.

In this work, we consider the above model  of FQHL tunnel junction
and find its quantum solution in the strong coupling limit. The zero frequency 
tunneling conductance obtained from this solution demonstrates three basic 
features. (1) The conductance depends on the assumed exchange statistics
between electrons in the different reservoir channels which appears in our solution
through the fermionic phase factor $e^{i\pi m}$ with in general arbitrary odd $m$.
(2) The standard choice of the minimum statistical phase $m=\pm 1$ results in the 
conductance $G_1=2 \sigma_0 \nu/(1+\nu)$ of a single tunneling mode. This result 
does not depend on the actual
number of the reservoir channels involved in tunneling. 
Such situation occurs \cite{me} in tunneling between the reservoir and the TLL wire
and, when the second reservoir is attached to the opposite end of the wire, results in the free electron conductance \cite{chkl,chamon} . 
(3) The "hyper-fermionic" \cite{hyper} statistics $m=1/\nu$ of the electrons induced by their
propagation along the edge leads to equilibration between the outgoing
edge and reservoir chemical potentials with increasing number of reservoir 
channels even in the absence of any decoherence  mechanisms \cite{kfcontact}. 

The appropriate choice of the statistics can be made
by minimization of energy or based on dynamics of charge propagation. 
A crossover occurs when the distance between the two tunneling 
point-contacts approaches an inverse momentum cut-off of the 1D edge excitations: 
For small separation between the contacts, the statistics is given by the minimum 
fermionic phase, and the tunneling is carried by a single tunneling mode, while 
for large separation the statistical phase is $1/\nu$ and the two reservoir 
channels can not be reduced to a single mode. A physical reason for the crossover 
is that an electron tunneling from the first channel into the edge at point $x_1$ and tunneling out at $x_2$ into the second channel propagates in the FQHL as a composite fermion \cite{jain} by absorbing  even number ($1/\nu-1$) of fluxes. Therefore, its correlation function $<\psi(x_2,i\tau)\psi^{\dagger}(x_1,0)> \propto (v \tau+i(x_2-x_1))^{-1/\nu}$ acquires a phase different from that of electrons in the chirally symmetric wire, where $<\psi(x_2,i\tau)\psi^{\dagger} (x_1,0)> \propto [(x_2-x_1)^2+(v\tau)^2]^{-(1/\nu-1)/2}/(v \tau+i(x_2-x_1))$. 
The large-time asymptotics of both functions are the same. Therefore, 
additional phase in the FQHL edge, which changes the hyper-fermionic extension of statistics of the tunneling electrons, depends on geometry of the junction.  
Technically, this phase is introduced through the commutation relations between  
bosonic fields appearing in tunneling operators at different points. The 
commutator does not vanish, since these fields are not chirally symmetric. 
This chiral asymmetry of tunneling operators  preserves time-reversal asymmetry 
of the quantum model even at low energies. 

The model we consider represents $n$ scattering channels of the spinless FL reservoir as free chiral fermions. Tunneling from the channels (labeled by 
$j=1,\, ...\,,n$) into the edge (labeled by $0$) is assumed to be localized 
on the scale of magnetic length at the points $x_j$ along the edge, where we 
take $x_j<x_i$ for $1 \le i<j \le n$. It is described by a tunnel Lagrangian:
\begin{equation}
{\cal L}_{tunn}= \sum_{j=1}^{n} [U_j\psi_0^+(x_j,t) \psi_j(x_j,t)+ h.c.]\, ,  
\label{H}
\end{equation} 
where $U_j$s are chosen real and positive. 
Bosonization expresses the operators of free electrons
$\psi_j(x,t)=(2\pi \alpha)^{-1/2} \xi_j e^{i\phi_j(x,t)}$ in the reservoir channels 
and the operator of electrons propagating along the edge 
$\psi_0=(2\pi \alpha)^{-1/2} \xi_0 e^{i\phi_0(x,t)/\sqrt{\nu}}$ 
through their associate bosonic fields $\phi_l$, the Majorana
fermions $\xi_l$ accounting for their mutual statistics, and a common
factor $1/\alpha$ denoting momentum cut-off of the edge excitations. Since the spatial dynamics of the reservoir channels does not affect the tunneling currents, velocities 
of these channels are irrelevant, and we take them as equal to the velocity $v$ of the edge excitations. Free dynamics of the bosonic fields is governed then by the Lagrangian ${\cal L}_O=\sum_{l=0}^n (\phi_l \hat{K}^{-1} \phi_l)/2$, where the differential 
operator $\hat{K}^{-1}$ is:
\begin{equation}
\hat{K}^{-1}\phi_l(x,t)={1 \over 2 \pi} 
\partial_x (\partial_t+ v \partial_x) \phi_l(x,t) \, .
\label{1}
\end{equation}
The full Lagrangian ${\cal L}={\cal L}_O+{\cal L}_\xi+{\cal L}_{tunn}$ also 
includes an additional kinematic part ${\cal L}_\xi=(1/4) \sum_l \xi_l \partial_t \xi_l$ describing a pure statistical dynamics of the Majorana 
fermions (time-ordering). 
A finite voltage applied to the reservoir is accounted for by the opposite-sign 
shift $\mu$ of the electrochemical potential of the edge. Since the chiral edge density $\rho_0$ is: $\rho_0= \sqrt{\nu} \partial_x \phi_0 /(2 \pi)$, this shift 
can be introduced by an additional non-equilibrium part of the Lagrangian 
${\cal L}_V=\sqrt{\nu}/2\pi \int dx \phi_0(x,t)\partial_x V(x)$, where 
$V(x)=-\mu \  \mbox{sgn}(x-y_X)$ and $y_X \to \infty$. With this choice of $V(x)$, 
the voltage introduced through ${\cal L}_V$ does not produce the edge current 
directly at $x<y_X$, and all the edge current $j_0=v \rho_0$ caused by ${\cal L}_V$ 
in the interval $x_1<x<y_X$ is just the opposite of the total tunneling current. 

To explain the main idea of our calculation we first consider the strong coupling 
limit of a one-point contact ($n=1$). The tunneling Lagrangian reduces to   
${\cal L}_{tunn}=U_1/(2\pi \alpha) \cos(\phi_0(x_1,t)/\sqrt{\nu}-\phi_1)$, 
and in the limit $U_1 \to \infty$ fixes the argument of the $\cos$-term at one of 
the cosine maxima, say, $\phi_0(x_1,t)/\sqrt{\nu}=\phi_1(x_1,t)$. 
This semiclassical procedure leads to an effective Lagrangian that is quadratic 
in its vector-argument $\phi(x,t) = [\phi_0,\phi_1]^T$. Performing functional 
integration over $\phi$ one can find the two-component average $\bar{\phi}(x,t)\equiv <\phi(x,t)>$ as 
\begin{equation}
\bar{\phi}(x,t)=-i {\sqrt{\nu} \over \pi} \int {d\omega \over \omega}
e^{-i \omega t-{\alpha \over v} |\omega|} g(x,\omega)\mu(\omega)\, ,
\label{average}
\end{equation} 
where $-2\pi i g/\omega$ is the first column of
the $(2 \times 2)$ matrix Green function. The two components $g_{0,1}(x)$ 
of the function $g$ do not depend on $y_X\to \infty$ for $x<y_X$, and 
satisfy the homogeneous differential equation (\ref{1}) at $x \not= x_1$ and 
therefore can be written as $g_{0,1}=a_{0,1}+b_{0,1} exp[-i \omega x/v]$.  
The coefficients take different values $a^\wp_{0,1},b^\wp_{0,1}$ 
for $x$ smaller  and larger than $x_1$ ($\wp$ denotes $<$ or $>$, respectively). 
They are related among themselves by four conditions: continuity of $g_{0}$ and $g_1$; continuity of the  current flow: $\sqrt{\nu} b^<_0+b^<_1=\sqrt{\nu} 
b^>_0+b^>_1$; and maximum of the tunneling term: $g_0(x_1)-\sqrt{\nu}g_1(x_1)=0$.  The solution $g$ is a linear combination of the four independent functions: 
$f^-_{c}=[\sqrt{\nu},1]^T, f^-_b=e^{i\omega x/v} f^-_c,
f^{<,>}_1=\theta(\mp(x-x_1))(e^{i\omega x/v}-1)[1,-\sqrt{\nu}]^T$, which are
constructed to satisfy these conditions.
Since propagation of tunneling electrons is governed by the diagonal free matrix Green 
function $K \times {\bf 1}$, where (for $\omega$  below cut-off): 
\begin{equation}
K(x-y,\omega)=-{2\pi i\over \omega} [{1 \over 2}+\theta(x-y)(e^{-i\omega(x-y) 
\over v}-1)],
\label{K}
\end{equation}
we can find more restrictions on the coefficients: $b^<_{0,1}=0, a^<_1=-a^>_1, a^\wp_0=1/2-a^\wp_1/\sqrt{\nu}$. They uniquely specify  $g(x,\omega)=[\sqrt{\nu}f^-_{c}/2 -f^>]/(1+\nu)$. The currents follow the from Eq.(\ref{average}) 
as $j_1(x,t)=-j_0= {2\nu \over 1+\nu} \theta(x-x_1) \sigma_0 \bar{\mu}(t-[x-x_1]/v)$,
where $\bar\mu(t)$ is the convolution of the function $\mu(t)$ 
with the normalized Lorentz factor of the width $1/D \equiv \alpha /v $. 
The tunneling conductance is  equal to $G_1=2\nu \sigma_0/(1+\nu)$ 
in agreement with the result of application \cite{chamon} of the chirally symmetric solution developed \cite{kf} for a point scatterer in TLL.

To extend this approach to the multi-channel contact, we notice that
although the statistical factors $\pm \xi_0 \xi_j$ attributed to 
annihilation/creation of electrons in the $j$th channel can not be ignored for 
more than one $j$ involved, they can be substituted \cite{JETPL} by the 
exponents $exp\{\pm i  \eta_j\}$ of the zero-energy bosonic fields
satisfying $[\eta_i,\eta_j]=i \pi \gamma_{i,j}$ with  odd integers $\gamma_{i,j}=-\gamma_{j,i}$. Different values of $\gamma_{i,j}$ correspond 
to different branches of the phase of the fermionic statistical factor $(-1)$ arising from interchange of $\xi_i$ and $\xi_j$. The pure statistical 
dynamics of the $\eta$-fields is described by the Lagrangian: 
${\cal L}_\eta=1/2\sum \eta_i \gamma'_{i,j} \partial_t \eta_j$, where 
$\gamma'_{i,j}$ are elements of the matrix inverse to the matrix 
$\gamma_{i,j}$. Substitution of the Majorana fermions by these bosonic 
modes transforms ${\cal L}_{tunn}$ into: 
\begin{equation}
{\cal L}_{tunn}\equiv \sum_{j=1}^n {\cal L}_j 
=\sum {U_j \over 2\pi \alpha} 
\cos\{{\phi_0(x_j,t) \over \sqrt{\nu}}-\phi_j-\eta_j\} \, . 
\label{3}
\end{equation} 
The models defined by Lagrangians ${\cal L}={\cal L}_O+{\cal L}_\eta+ {\cal L}_{tunn}$ with different $\gamma_{i,j}$ are equivalent. This can be seen from
comparison of their perturbative expansions in ${\cal L}_{tunn}$.
Indeed, since any non-vanishing term of these expansions contains $\pm$ 
exponents in pairs, a proper interchange of the exponents cancels them all 
and leaves only the statistical sign. The sign is the same as one would get 
directly from the Majorana fermions and does not depend on the choice of 
matrix $\gamma$. When all $U_j$ in (\ref{3}) become large, 
the arguments of all the cosine functions take the values that maximize 
them. The semiclassical condition that these arguments can be simultaneously 
fixed, imposes  restriction on the phase space of the system. This restriction  
depends on the choice of matrix $\gamma$, and the physically relevant $\gamma$ 
can be determined by minimizing the energy contribution produced by this restriction. In the case of the two-point junction, there is only one parameter 
$\gamma\equiv \gamma_{2,1}$, and this can be done explicitly by using 
imaginary time representation of the Lagrangians, and evaluating \cite{future} 
two parts of this contribution coming from modes with the energies above and below $T_{12}=v/(x_1-x_2)$ as $(D-T_{12})ln[(1+\nu)^2+(\gamma\nu-1)^2]$ and $T_{12} ln[(1+\nu)^2+\gamma^2\nu^2-1]$, respectively. The choice $\gamma=1/\nu$ that makes the same-time commutator of the two cosine arguments in (\ref{3}) vanish, minimizes the high-energy contribution. This means that this value of $\gamma$ has to be used if the two tunneling points are well separated so that $T_{12} \ll D$. When, however, two tunneling points approach each other and $T_{12} \simeq D$, the loow-energy contribution becomes more important, and the 
choice of $\gamma$ has to be changed to the smallest value consistent with the  statistics ($\gamma=1$ in our case).  

We can use the strong-coupling conditions to perform real-time calculation of 
the current flow in a multi-point contact.  The real-time results provide another, more intuitive, interpretation of the choice of  the statistical phases. We parametrize them as 
 $\gamma_{i,j}=\gamma \mbox{sgn}(i-j)$, where $\gamma$ is kept as a free parameter. 
This form is sufficiently general to describe the two-point contacts or contacts where all 
tunneling points are  well-separated from one another.  The strong-coupling conditions $\phi_0(x_j,t)/ \sqrt{\nu}-\phi_j-\eta_j=0,\ j=1 \div n$ make Lagrangian of the model Gaussian,  and linear relations between the currents and the bias voltage are determined by the retarded Green 
function. The calculation generalizes the one for the single-point contact. The average $\bar{\phi}(x,t)$ in Eq. (\ref{average}) becomes the $(n+1)$-component vector 
$<[\phi_0,...,\phi_j+\eta_j,...]^T>$, and $-2\pi i g/\omega$ is 
the first column of the $(n+1) \times (n+1)$ matrix Green function. 
The coefficients $a_j, b_j, j=1 \div n$ take different 
values $a^\wp_j, b^\wp_j$ for $x$ smaller and larger than $x_j$, where $\wp$ denotes $<$ and $>$ as before. The edge channel coefficients $a_0,\, 
b_0$ take $(n+1)$ different values, changing at each tunneling contact $x=x_j$ in a way that relates them to $a_j, b_j$ by the four matching conditions derived above for the single-contact case. We denote with $a_0^\wp$ and $b_0^\wp$ their values for $x$ smaller than $x_n$ ($\wp=\, <$) and larger than $x_1$ ($\wp=\, >$). A set of $2(n+1)$ independent vector-functions satisfying all these conditions can be chosen as: $f^-_c=[\sqrt{\nu},1,1,1,...]^T , f^-_b=e^{i\omega x/v} f^-_c,
f_j=(e^{i\omega x/v}-e^{i\omega x_j/v}) e_j,
f^>_j=(e_0/\sqrt{\nu}-e_j)\theta(x-x_j)(e^{i\omega (x-x_j)/v}-1)+ 
\sum_{l=1}^{j-1}e_l (e^{i\omega (x_l-x_j)/v}-1)/\nu $, where  a vector $e_l$ has the
only non-zero $l$th component equal to $1$. Since all $b^<_l$ coefficients of the function $g$ are zero, it can be expanded in this basis as 
\begin{equation}
g= s_c f^-_c + \sum_{j=1}^n s_j f^>_j   
\label{g}
\end{equation} 
with $(n+1)$ unknown coefficients $s_l$. The non-zero $s_j$ lead to finite jumps 
of $a_j$ and $b_j$ at $x=x_j$, and therefore, to the non-vanishing $b^>_j=-s_j e^{-i\omega x_j/v}$. Then, in accordance with Eq.\ (\ref{average}), the reservoir channel currents arising at $x_j$ can be found as $j_j(\omega,x)= 2 \sqrt{\nu} \mu \sigma_0 b^>_j \theta(x-x_j) e^{i\omega x/v}$. Jumps of the coefficients $a_j$ and $b_j$ are caused by the charge tunneling at the 
contact points $x_j$, with further propagation of charge after tunneling governed 
by the free retarded Green function. This means that this function determines both 
the continuous parts of the $a,\ b$ coefficients and the relations between their discontinuous parts and the coefficients $s_j$. The Green function is $(n+1) \times (n+1)$ matrix, and can be written as $K \times {\bf 1} - \gamma \pi i  {\bf 
C}/\omega$, where $\bf C$ is the antisymmetric matrix with all elements above the diagonal, except the first row, equal to $1$. From this form one can find that $b_l^<=0$ (the fact already used in Eq.\ (\ref{g})), and that the coefficients 
$a_l$ are related to $s_j$. In particular: $a^<_0=1/2+\sum_{p=1}^n s_p /2\sqrt{\nu},
a^<_j=-s_j/2+\gamma/2 \sum_{p \neq j} \mbox{sgn}(j-p) s_p$. From comparison of these 
relations with those obtained by direct substitution of the $f$-vectors into Eq. (\ref{g}) we find $n$ equations: 
\begin{eqnarray}
{\sqrt{\nu} \over 1+\nu}&+&s_n=-{1-\gamma \nu \over 1+ \nu} \sum_1^{n-1} s_i
\label{s}\, , \\
s_p=s_n &+&  \sum_{j=p+1}^n
{2 s_j \over (1+\gamma)} [(1-e^{i\omega(x_p-x_j)/v})/\nu-\gamma]\, ,
\nonumber
\end{eqnarray}
where $p=1 \div (n-1)$. Equations (\ref{s}) allow us to determine all unknown  coefficients $s_j$.

For the two-point contact these equations reduce to:
\begin{eqnarray}
s_1&=&(1-\gamma+{2 \over \nu}[1-e^{i\omega (x_1-x_2)/v}]){s_2 \over 1+\gamma} 
\, , \nonumber \\
s_2&=&-{\nu^{3/2}(1+\gamma)\over 2R}\, ,
\label{c2} \\
R&\equiv &1+\nu(1-\gamma)+{\nu^2 \over 2}(1+\gamma^2)
-(1-\nu \gamma) e^{i\omega(x_1-x_2) \over v} \nonumber \, .
\end{eqnarray}
The part of the denominator $R$ proportional to $(1-\nu \gamma)$ 
signals the appearance of an interference structure in the currents.  
Substituting $s_{1,2}$ (\ref{c2}) into 
$j_0(x,t)=-\int d\omega e^{-i\omega t-\alpha |\omega|}\sum_{1,2}j_j(\omega,x)$ 
one can see that, indeed, the time dependence of 
charge propagation along the edge exhibits multiple backscattering events at 
$x_2$ and $x_1$: A charge wave started by the tunneling into the edge  
propagates from the point $x_2$ to $x_1$ with the velocity $v$ and then instantly recoils back to $x_2$ from $x_1$ with a finite reflection coefficient 
proportional to $(1-\nu \gamma)$. 
The formal possibility of the charge propagation with infinite velocity in the direction opposite to the edge chirality is a combined effect of 
$x$-independent solutions of the operator $\hat{K}^{-1}$ from (\ref{1}) and the
matching conditions at the tunneling points. However, this instant ``counter-propagation'' violates causality of the edge response to external perturbations and can not appear in the final physical results. 
This argument, in agreement  with the minimization of the energy contributions, makes 
$\gamma=1/\nu$ the only possible choice in the situation when $x_1$ and $x_2$ are well separated 
and $T_{12} \ll D$. When $x_1$ is so close to $x_2$ that $T_{12}$ is on the order of $D$, the choice
$\gamma=1$ does not contradict the causality, since charge density $j_0(x,t)$ can not be split into two different parts located at these points.  

A general expression for the zero-frequency tunneling conductance $G$ valid for any 
value of $\gamma$ and number of the reservoir channels $n$ can be obtained  from (\ref{s}). The zero-frequency solution of the second of Eqs. (\ref{s}) is: $s_{p-1}=q s_p, p=2 \div n$, with $q=(1-\gamma)/(1+\gamma)$ and makes all $s_j$  proportional to $s_n$. The conductance $G=-2 \sigma_0 \sqrt{\nu} \sum s_j$ is obtained then 
from the first of Eqs.\ (\ref{s}):
\begin{equation}
G=\frac{2 \nu [(1+\gamma)^n-(1-\gamma)^n]}
{(1+ \nu \gamma)(1+\gamma)^n-(1-\nu \gamma)(1-\gamma)^n} .
\label{G}
\end{equation}
This expression shows how the conductance depends on the assumed exchange statistics between electrons in the different reservoir channels. The choice
$\gamma=1$ corresponds to the tunneling model where all point contacts are located
closely to each other and for any $n$ yields $G=G_1$, i.e., all reservoir channels are reduced to one tunneling mode. 
After incorporation of the similar tunneling junction with the second FL reservoir,
the one-tunneling-mode conductance becomes
equal  \cite{me,chkl} to the free electron conductance $\sigma_0$.
If all point contacts are well separated we should use $\gamma=1/\nu$. 
The tunneling conductance is then $ G=\sigma_0 \nu (1-q^n)$ and 
depends on the number of channels. 
When $n \to \infty$, it saturates at $ \nu \sigma_0$. 
All currents calculated in this regime are temperature independent and coincide with those derived under the assumption of decoherence \cite{chamon,kfcontact}. 
To our knowledge, this assumption does not have quantitative microscopic justification. In particular, due to chiral propagation along the edge, a large temperature by itself \emph{does not} lead to decoherence \cite{future}. As can be seen from the solution constructed above, the same incoherent addition of currents can be obtained without decoherence, when separation of contacts leading to equilibration inside the edge is achieved by appropriate choice of statistical phases. 

As a final point, we use the constructed strong coupling solution to discuss 
fluctuations of the tunnel currents in the multi-point junction. The  
fluctuations are characterized by the symmetric correlator of pairs of
the fluctuating parts of the currents $j_i(x,t)-<j_i(x,t)>, \, i=0\div n$.  
As before, when positions of the tunneling contacts allow a definite choice of factors  
$\gamma$, the phases in (\ref{3}) are fixed in the strong coupling regime, and 
the Lagrangian acquires a Gaussian form. In this case, effect of the finite bias voltage in the Lagrangian is reduced to the shift of the $\phi$ averages which 
cancel out in the correlators. Therefore, the junction in this regime 
does not generate the shot noise, i.e., even at finite bias voltages the current fluctuations are the same as in equilibrium, and at zero frequency vanish linearly with temperature. 

In conclusion,  we have found the strong-coupling solution of the model of  tunneling between the multi-channel Fermi-liquid reservoir and an edge of the principal FQHL. Solution  depends on the choice of the statistical phase branches of the different channels.  When the reservoir channels tunnel into the edge at well-separated points, the phase branches are uniquely determined by the requirement of conservation of the initial commutation symmetry of the appropriate parts of the tunneling Lagrangian. Additional statistical phases can be interpreted as even number of fluxes absorbed/emitted by electrons tunneling  into the edge that transform them into composite fermions. As tunneling points approach one another, the regular statistical phases  $\pm\pi$ are restored and the multi-channel tunneling is reduced to a single tunneling mode. Our results  explain the difference between transport through the 1D FQHL edge and the Tomonaga-Luttinger-liquid wire: The two terminal universal conductance of the edge is renormalized by the flux attachment, while direct electron-electron interaction in the wire does not change its universal 
free-electron conductance.

V.P. acknowledges fruitful discussions with N. Nagaosa and M. Buttiker on 
the preliminary stage of this work. This work was supported by the NSA and 
ARDA under the ARO contract.

\end{document}